\documentclass[format=acmsmall, review=false, screen=true]{acmart}

\usepackage{amsmath,amssymb,amsfonts}
\usepackage{algorithmic}
\usepackage{graphicx}
\usepackage{textcomp}

\usepackage{pgfplots}
\usepackage{hyperref}
\usepackage{bm}
\usepackage{multirow}
\usepackage{booktabs} 

\usepackage{pgfplotstable}
\pgfplotsset{compat=newest}
\usepackage{caption}
\usepackage{subcaption}

\usepackage{mathtools}
\usepackage{stackengine}
\usepackage{easybmat}
\usepackage{tikz}
\usepackage{multirow,bigdelim}

\definecolor{dgreen}{HTML}{219601}
\definecolor{bblue}{HTML}{4F81BD}
\definecolor{rred}{HTML}{C0504D}
\definecolor{oorange}{HTML}{ffa435}
\definecolor{ggreen}{HTML}{9BBB59}
\definecolor{dred}{HTML}{5c0802}
\definecolor{bblue}{HTML}{15b9ff}

\newcommand*\hexbrace[2]{%
  \underset{#2}{\underbrace{\rule{#1}{0pt}}}}

\setcopyright{acmlicensed}

\renewcommand\footnotetextcopyrightpermission[1]{} 
\usepackage{fancyhdr}
\acmDOI{0000001.0000001}

\begin{document}

\title{Exploring Users' Perception of Collaborative Explanation Styles}

\author{Ludovik Coba}
\affiliation{%
	\institution{Free University of Bozen}
	\streetaddress{Piazza Domenicani 3}
	\city{Bolzano}
	\state{39100}
	\country{Italy}
}
\email{ludovik.coba@inf.unibz.it}

\author{Markus Zanker}
\affiliation{%
	\institution{Free University of Bozen}
	\streetaddress{Piazza Domenicani 3}
	\city{Bolzano}
	\state{39100}
	\country{Italy}
}
\email{Markus.Zanker@unibz.it}

\author{Laurens Rook}
\affiliation{%
	\institution{Delft University of Technology}
	\city{Delft}
	\country{The Netherlands}
}
\email{l.rook@tudelft.nl}

\author{Panagiotis Symeonidis}
\affiliation{%
	\institution{Free University of Bozen}
	\streetaddress{Piazza Domenicani 3}
	\city{Bolzano}
	\state{39100}
	\country{Italy}
}
\email{Panagiotis.Symeonidis@unibz.it}

\begin{abstract}

Collaborative filtering systems heavily depend on user feedback expressed in product ratings to select and rank items to recommend. In this study we explore how users value different collaborative explanation styles following the user-based or item-based paradigm. Furthermore, we explore how the characteristics of these rating summarizations, like the total number of ratings and the mean rating value, influence the decisions of online users. Results, based on a choice-based conjoint experimental design, show that the mean indicator has a higher impact compared to the total number of ratings. Finally, we discuss how these empirical results can serve as an input to developing algorithms that foster items with a, consequently, higher probability of choice based on their rating summarizations or their {\it explainability} due to these ratings when ranking recommendations.

\end{abstract}



\maketitle

\section{Introduction}

User ratings are one of the key ingredient to collaborative filtering algorithms to automatically assess how likely items might match users' tastes. 

Although, recently, implicit signals on users' actual behavior have turned out to possess even more predictive power for practical systems \cite{gomez2016netflix}, ratings still play a dominant role in constructing the value and quality perception of an item in the eyes of online consumers \cite{duan2008online}. 

Collaborative explanations \cite{Friedrich2011ASystems} provide justifications for recommendations by displaying information about the rating behavior of a users' or items' neighborhood, as has been already identified by Herlocker et al. \cite{Herlocker2000ExplainingRecommendations}. Also, with the products in their catalogs, e-commerce sites usually provide at least rating summary statistics along with an information about the origin the ratings.

In this paper we therefore present a choice-based conjoint study that investigates two aspects of these collaborative explanations. The {\it first aspect} regards the users' perception of three different origins for collaborative rating summarizations, i.e.:
\begin{itemize}
\item summaries derived from ratings of users with similar online behavior to the current user in terms of ratings, purchases or clicks (user-style explanations), 
\item summarizations based on the ratings from the social-network friends of the current user (social explanations), and
\item ratings of the current user given to similar items, e.g., {\it this is how you rated similar movies to this one} (item-style explanations).
\end{itemize}
The {\it second aspect} of our study relates to the two dominant characteristics of a rating summarization, namely number of ratings and mean value, and how they impact the choice behavior of users.  
When investigating preferences for the origin of ratings, our results show that users clearly prefer rating summarizations justified by similar users (user-style explanations) or similar items (item-style explanations), over ratings from social network friends.
While results on the characteristics of the ratings summarizations show that - all things being equal - users are clearly biased towards selecting items with higher means as opposed to larger numbers of ratings. 
Thus, this study provides clear indications about the degree of {\it persuasiveness} \cite{yoo2012persuasive} of these different aspects of collaborative explanations.  

After outlining related work in Section~\ref{sec:rel_work}, we will give details on the choice-based conjoint methodology used for performing the user study in Section~\ref{sec:cbd_design}. In Section~\ref{sec:results} we outline obtained results and finally, in Section~\ref{sec:discussion}, discuss implications for recommender systems research.

\section{Related work}
\label{sec:rel_work}
\begin{figure}
\centering
\fbox{\includegraphics[scale=0.55]{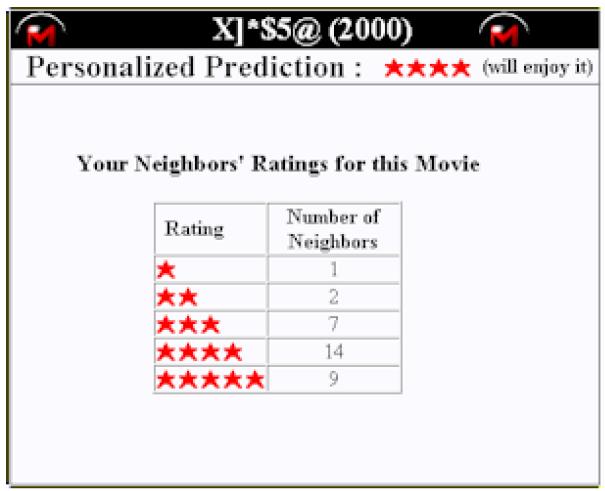}}
\caption{Collaborative user-style explanation from \cite{Herlocker2000ExplainingRecommendations}.}
\label{fig:herlocker}
\end{figure}

\begin{figure}
\centering
\fbox{\includegraphics[scale=.6]{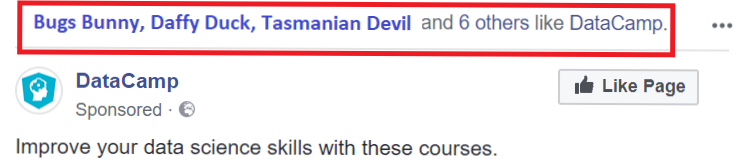}}
\caption{Example of justification using Facebook friends.}
\label{fig:fbstyle}
\end{figure}

Explanations for recommendations have received considerable research attention over the past years, as summarized by \cite{Tintarev2015ExplainingEvaluation} and \cite{Nunes2017ASystems}. There are different ways of explaining recommendations based on collaborative filtering mechanisms as presented in Herlocker et al.~\cite{Herlocker2000ExplainingRecommendations}. They explored 21 different interfaces and demonstrated that specifically the ``user" style (see Figure~\ref{fig:herlocker}) improves the acceptance of recommendations. 
The ``user" style of explanation provides information about the neighborhood, which is determined based on a generic notion of similarity between users when analyzing their observed behavior or expressed opinions (i.e., buys, clicks, ratings etc.). Please notice that social links (e.g. Facebook friends, see Figure~\ref{fig:fbstyle}) can be considered as a special case of the user style of justifications \cite{Papadimitriou2012ASystems}.
As far as the user style of explanation is concerned, several collaborative filtering recommender systems, such as Amazon, adopted the following style of justification: ``Customers who bought item $X$ also bought items $Y, Z, \ldots$". 

\begin{figure}
\centering
\includegraphics[scale=.5]{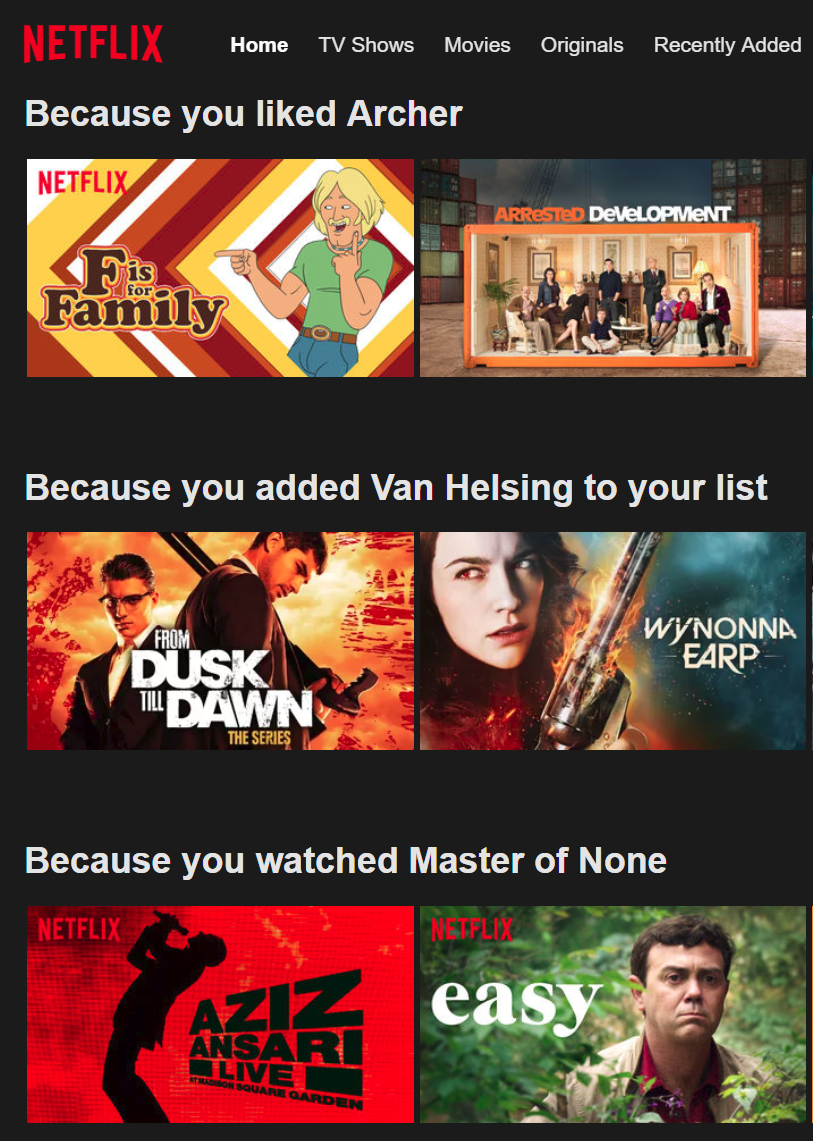}
\caption{Items style explanation example from the Netflix system.}
\label{fig:itemStyleN}
\end{figure}

In the so-called item style of explanation, the justifications are of the following form: ``Item $Y$ is recommended because you highly rated/bought item $X, Z, \ldots$". Thus, the system depicts those items i.e., $X, Z, \ldots$, that influenced the recommendation of item $Y$ the most. Bilgic et al.~\cite{Bilgic2005ExplainingPromotion} claimed that the item style is better than the user style, because it allows users to accurately formulate their true opinion of an item.

Several works researched the effectiveness of this explanation strategy~\cite{Cosley2003,Bilgic2005ExplainingPromotion,Papadimitriou2012ASystems}. Rating summary statistics have become common patterns to explain recommendations in many domains\cite{Cremonesi2017UserApplications}.

In this line of research, Cosley et al. \cite{Cosley2003} noticed, for instance, that presenting fine grained rating information in a recommendation is highly desirable. However it might bias the users' opinion,i.e. promote items as opposed to increasing the effectiveness.

In contrast to the aforementioned works, however, we are interested in shedding light on users' trade-off between rating numbers and their mean values when they have to make a choice. 
 
Conjoint analysis is a market research technique suitable for revealing user preferences and trade-offs in the decision making process\cite{Rao2014ChoiceAnalysis}. Conjoint analysis has successfully been employed in a wide range of areas, such as education, health, tourism, and human computer interaction.

Cho et al.\cite{Cho2015TheStyle} conducted a conjoint experiment to investigate elders' preference over smart-phone application icons. The authors explored the dynamics of two attributes (degree of realism and level abstraction) one with four levels and one with two levels, and ran their user study with a total of 30 respondents.

In the field of recommender systems and online decision support, Zanker and Schoberegger \cite{Zanker2014AnSystems} employed a ranking-based conjoint experiment to understand the persuasive power of different explanation styles over the users' preferences. 
More recently, Carbonell et al. \cite{Carbonell2018ChoosingPhysician} observed that users select physicians based on considerations of user generated content such as ratings and comments rather than the official descriptions of the physicians' qualifications. The authors used a choice-based conjoint design to understand, which features influenced the users choice, and suggested that including these results in recommender systems would improve the decision making process.

However, to the best of our knowledge, the persuasive effect of the characteristics in rating summarizations has not yet been studied. The conjoint methodology as employed in market research for decades represents a best practice in order to quantify the perceived utility of the characteristics of different rating summarizations.

\section{Methodology and design}
\label{sec:cbd_design}
\textit{Choice-based Conjoint} (CBC) analysis is a frequently used approach to determine users' preferences over a wide range of attributes characterizing products or services~\cite{Chu2009AYahoo,Kuhfeld2010DiscreteChoice}. The Choice-Based Conjoint (CBC) methodology is also denoted as Discrete Choice Experiment by several authors \cite{louviere2010discrete}. 
In this Section, we explain the used approach in investigating the user's perception of the rating summarizations and explain how we developed and deployed the CBC questionnaire. 

The study is divided into two tasks, one designed to investigate how users perceive different origins of ratings, and the other to investigate the trade-off mechanisms between different characteristics of rating summarizations. We had two separate designs of the experiment in order to consider and test attribute levels that are representative for both, the item-style and the user-style of explanations in the movie domain.

\subsection{Acceptance of the origin of ratings}

We measured users' preference for three different origins of ratings summarizations, two variations of the user-style of explanations (i.e. similar users and friends on social networks), and the item-style of explanations (i.e. user's ratings on similar items). 
We designed three profiles, each introduced with one of the sentences presented in Table~\ref{tab:origin}, and followed by the identical rating summarization, thus only the origin of the ratings summarized below differed.

\begin{figure}
\centering
\includegraphics[scale = 0.9]{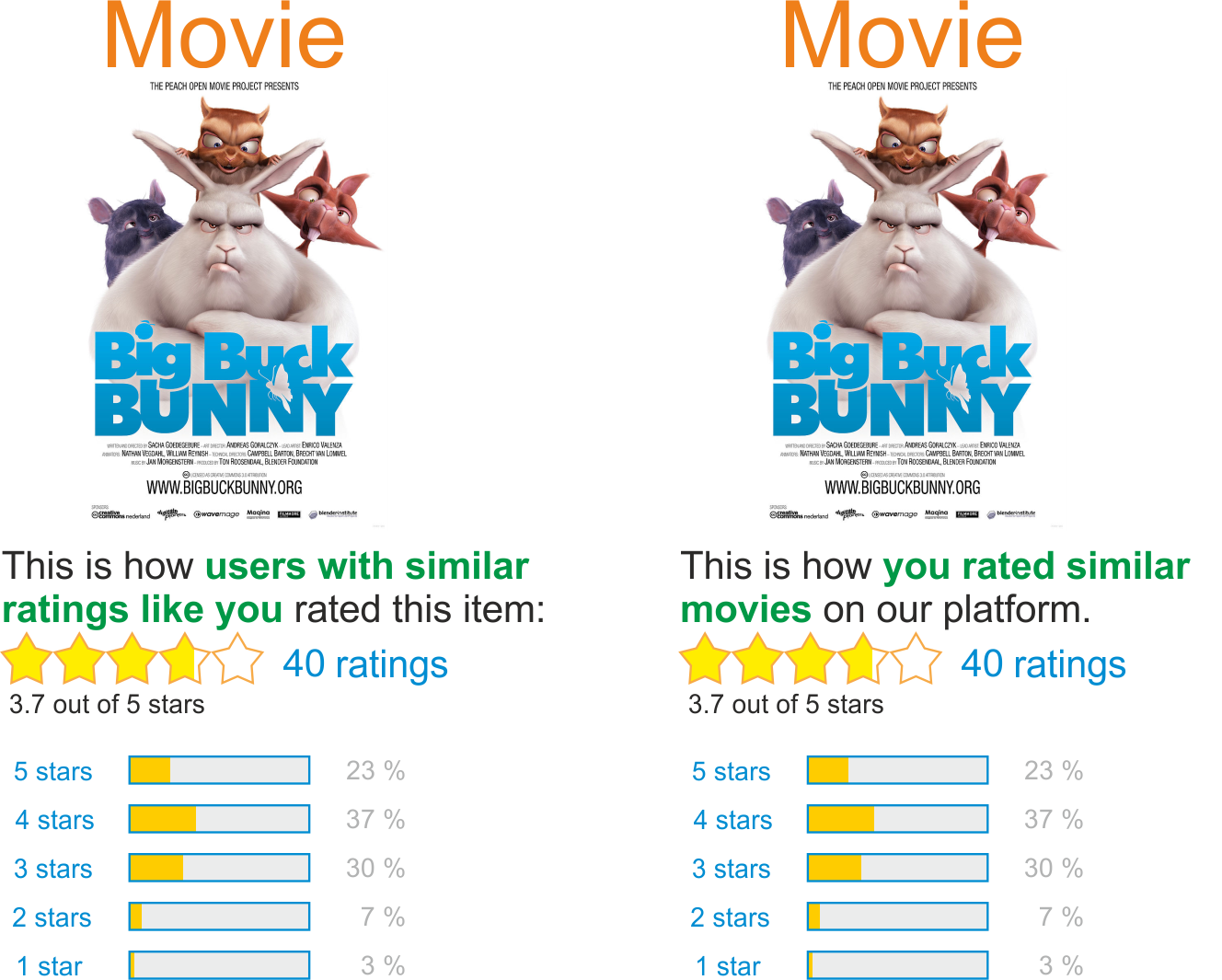}
\caption{Example of choice between two different origins of ratings. The movie poster was adopted from: \url{https://peach.blender.org/}.}
\label{fig:origin_of_ratings}
\end{figure}

Users had three binary choices between two out of the three different categories of origins of the ratings, like depicted in Figure \ref{fig:origin_of_ratings}.

For this task respondents were confronted with the following choice scenario:

\textit{``Assume that you find yourself in the situation that you want to make a choice between two different movies to watch.
Furthermore, assume that you only care about the origin of the ratings that are presented for each movie, i.e. ratings from other users that had similar preferences like you in the past, ratings of your friends on Facebook or your own ratings for movies that are similar to the one you look at. 
We therefore would like to ask you about your preference if solely based on this origin of the ratings.''}

\begin{table}
\caption{The stimuli presented in the origin of ratings experiment.}
\begin{tabular}{cl}
 \hline
 & Origin of Ratings\\
 \hline
 1 & This is \textbf{how users with similar ratings like you} rated this item \\
 2 & This is how \textbf{your friend on Facebook} rated this item\\
 3 & This is how \textbf{you rated similar movies} on our platform\\
 
 \hline
 
\end{tabular}
\label{tab:origin}
\end{table}

This task round was completed with a \textit{manipulation check} to validate  respondents' correct perception of our stimuli. 
In the manipulation check, we asked participants about the strategy they had employed in the making of their choices. 
Based on their answers, we only included those participants who reportedly had noticed that the origin of the summarized ratings differed between choices like ratings from similar users or on similar items. Thus, we removed those respondents who reportedly solely relied on their gut feeling for making their decision.

\subsection{Choice-Based Conjoint (CBC) methodology}
 
By collecting answers from different choice sets, researchers can quantify the impact of an attribute level on the preference of respondents~\cite{Hauber2016StatisticalForce}. 
In conjoint designs, products (a.k.a., {\it profiles}) are modeled by sets of categorical or quantitative \textit{attributes}, which can have different \textit{levels}. In CBC experiments, participants have to repeatedly select one profile from different \textit{sets of choices}, which nicely matches real-world settings when users are confronted with recommendation lists.

\subsubsection{Selection of attributes}
\label{selection_ofAttributes}

\begin{figure}
\centering
\includegraphics[scale=1]{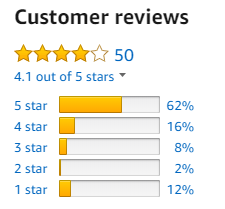}
\caption{Example plot of a ratings summarization from Amazon.com. Their model is not just a raw data average of the reviews but also considers factors such as the age of the review.}
\label{fig:amazon_example}
\end{figure}

The first step in building a conjoint design is determining the attributes and their corresponding levels. The most striking characteristics of rating summarizations (see, for instance, Figure~\ref{fig:amazon_example}) are the \textit{number of ratings} and the \textit{mean rating value} selected as attributes in our conjoint choice design.

The total {\it number of ratings} is often seen as a proxy to measure an item's popularity, and many well known algorithms are implemented to recommend items that are frequently rated~\cite{Jannach2015}. 
Following the argument of ~\cite{deLanghe2016NavigatingRatings} a big number of ratings with a slightly lower rating mean should be preferred over higher means based on a much lower total number of ratings. This leads us to the second attribute of this study, the {\it mean rating value}. 
Formally, a rating summary statistic is a frequency distribution on the class of discrete rating values. Thus, besides the total number of ratings and the mean, also the factors variance and skewness are needed for an approximate description of a unimodal rating distribution \footnote{Empirically, one can also observe bimodal rating distributions as depicted, for instance, in Figure \ref{fig:amazon_example}}. In our design, we controlled for variance and skewness  of rating distributions by keeping them fixed. In order to ensure a representative choice of attribute levels for the movie domain, we relied on the Netflix dataset (see Table~\ref{tab:netflixDs}) to identify real-world levels for characterizing rating frequency distributions. Note, that the Netflix dataset itself is not needed to reproduce our study, but only the attribute levels derived from the dataset as described in this paper. The Netflix dataset consists of 17,770 items, 480,189 users and contains 100,480,507 ratings on a discrete scale ranging from 1 to 5.  It has been heavily used in recommender systems research and provides evidence for the relatively high number of ratings on movie items. 

\begin{table}
\caption{Netflix datasets.}
\label{tab:netflixDs}
\centering
\begin{tabular}{lc}
\toprule
Number of ratings & 100,480,507\\
Rating's domain & [1;5]\\
Mean rating value & 3.6 \\
\# of items & 17,770 \\
Average \# of ratings per item & 5654.5\\
\# of users & 480,189 \\
Average \# of ratings per user & 209.3\\
\bottomrule
\end{tabular}

\end{table}

\begin{figure*}
\centering
\begin{subfigure}[b]{0.45\textwidth}
\includegraphics[scale=0.7]{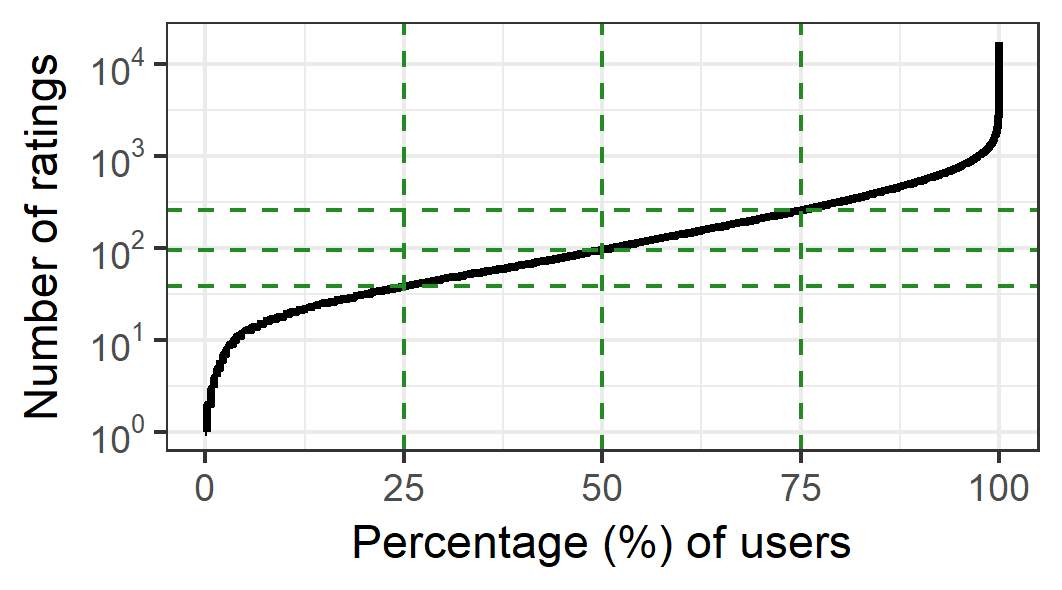}
        \caption{}
        \label{fig:user_ratings}
\end{subfigure}
\begin{subfigure}[b]{0.45\textwidth}
\includegraphics[scale=0.7]{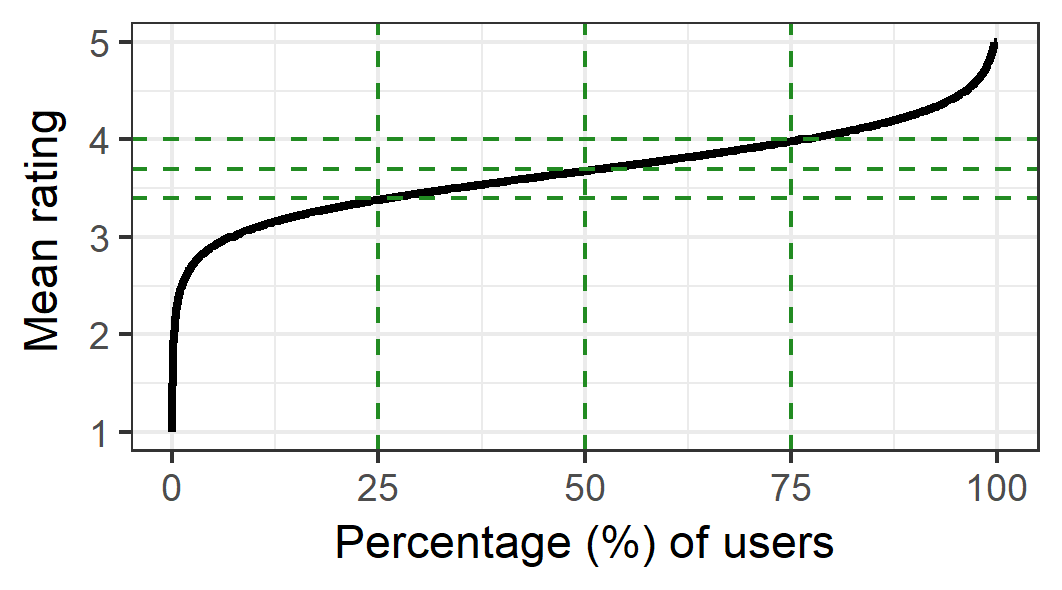}
        \caption{}
        \label{fig:user_mean}
\end{subfigure}
\caption{Rank distribution of users based on the (a) number of ratings and (b) mean value, in the Netflix dataset.}
\label{plot:user_stat}
\end{figure*}

\begin{figure*}
\centering
\begin{subfigure}[b]{0.45\textwidth}
\includegraphics[scale=0.7]{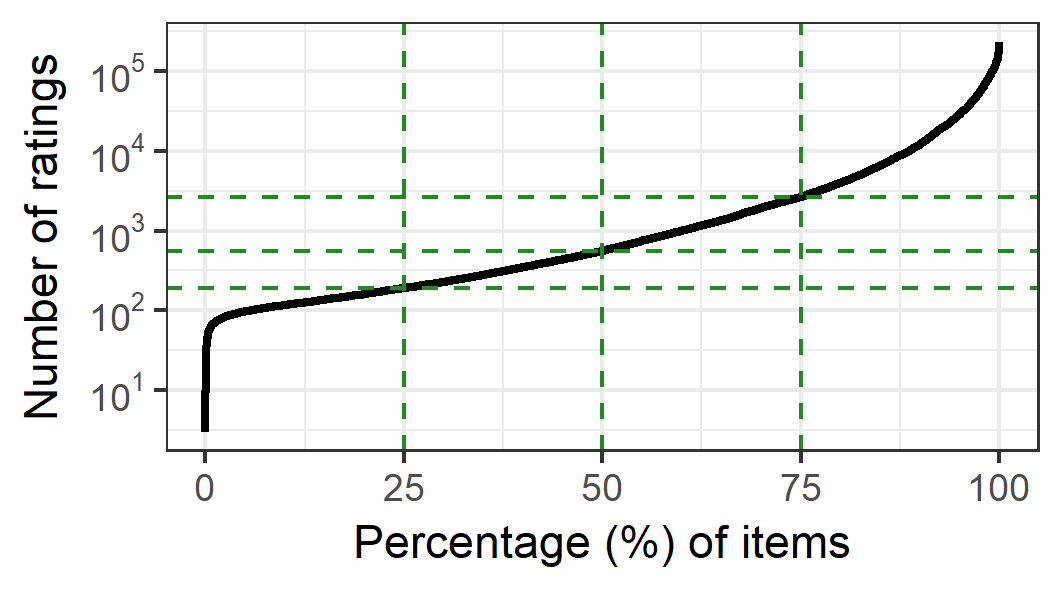}
        \caption{}
        \label{fig:item_ratings}
\end{subfigure}
\begin{subfigure}[b]{0.45\textwidth}
\includegraphics[scale=0.7]{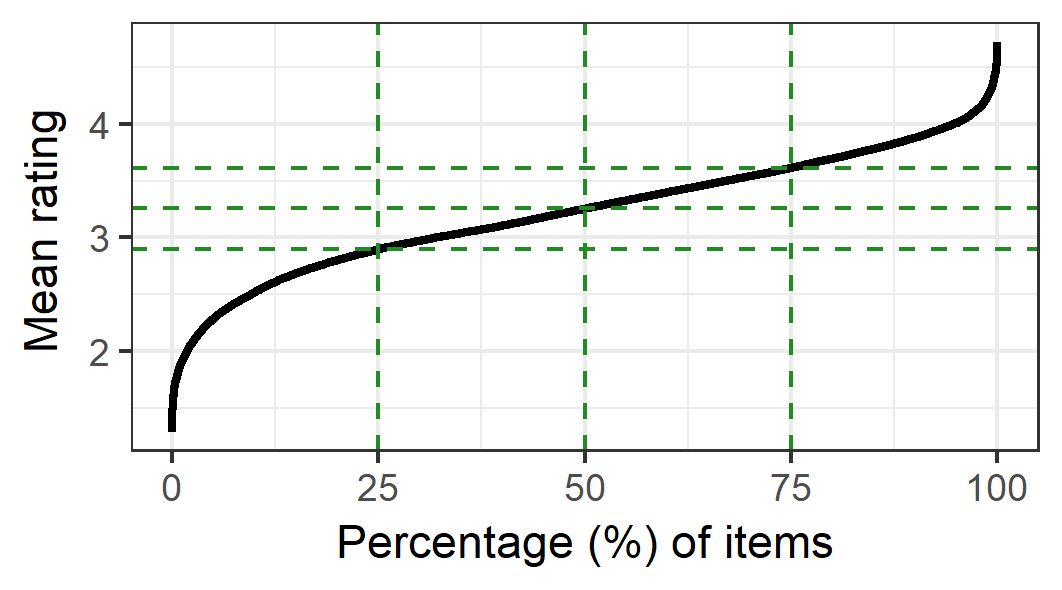}
        \caption{}
        \label{fig:item_mean}
\end{subfigure}
\caption{Rank distribution of users based on the (a) number of ratings and (b) mean value, in the Netflix dataset.}
\label{plot:item_stat}
\end{figure*}

\begin{table}[h]
\centering
\caption{Attributes and attribute levels in the Ratings values experiment.}
\label{table:attributes_levels}
\begin{tabular}{lll}
\hline
\multirow{2}{*}{Attribute} & \multicolumn{2}{c}{Levels}  \\
\cline{2-3}
& Item based & User based\\
\hline
\multirow{3}{*}{A1: Number of Ratings} 	& L1: 39 (small) & L1: 290 (small) \\
						& L2: 96 (medium)& L2: 560 (medium) \\
                        & L3: 259 (big) & L3: 2970 (big)\\
                        \cline{2-3}
\multirow{3}{*}{A2: Mean Rating	}		& L1: 3.4 (low) & L1: 2.9 (low)\\
						& L2: 3.7 (average) & L2: 3.3 (average) \\
                        & L3: 4 (high) & L3: 3.6 (high)\\
 \hline

\end{tabular}
\end{table}

User-style rating summarization depends on ratings given by other users on the same item, while item-style summarizes the ratings for similar items of the current user. We therefore opted for two different level combinations for number of ratings and mean attributes that we tested on two different samples of participants.

In order to determine the attribute levels for the item-style, we analyzed the distribution of ratings per user in the Netflix movie dataset. Figure~\ref{fig:user_ratings} shows the rank distribution of the users based on the total number of ratings. The 25th, 50th and 75th percentiles (i.e., lower quartile, median and upper quartile) of the number of ratings are 39, 96, and 259, which we, henceforth, denote as the {\it Small}, {\it Medium} and {\it Large} condition for the number of ratings. Next, Figure~\ref{fig:user_mean} analogously depicts the rank distribution of the mean rating values. The 25th, 50th and 75th percentiles have rounded mean rating values of 3.4, 3.7, and 4, respectively, which are our {\it Low}, {\it Average} and {\it High} conditions for the mean rating values.

While, for the run on the user-style we determined the levels by analyzing the rating distributions per item. Figure~\ref{fig:item_mean} shows the rank distribution of the users based on the total number of ratings. Again, the 25th, 50th and 75th percentiles of the number of ratings are 290, 560, and 2970 (denoted as the {\it Small}, {\it Medium} and {\it Large} conditions for the number of ratings). Note, that these levels are several times bigger than for the item-style, since obviously items attract high numbers of ratings in the movie domain. Next, Figure~\ref{fig:user_mean} analogously depicts the rank distribution of the mean rating values. As before, the 25th, 50th and 75th percentiles have rounded mean rating values of 2.9, 3.3, and 3.6 ({\it Low}, {\it Average} and {\it High} conditions for the mean rating values), smaller than analog conditions in the items based, explainable by the higher density of interactions per item rather than per user.

Table~\ref{table:attributes_levels} summarizes the selected attributes and the selected values for each level.
In addition, we controlled for variance and skewness of the rating frequency distributions by fixing them with the median values from the respective Netflix rank distributions for both runs of the user study (variance: 1 and skewness: -.5).

\subsubsection{Study design}

\begin{figure*}
\centering
\includegraphics[scale=0.9]{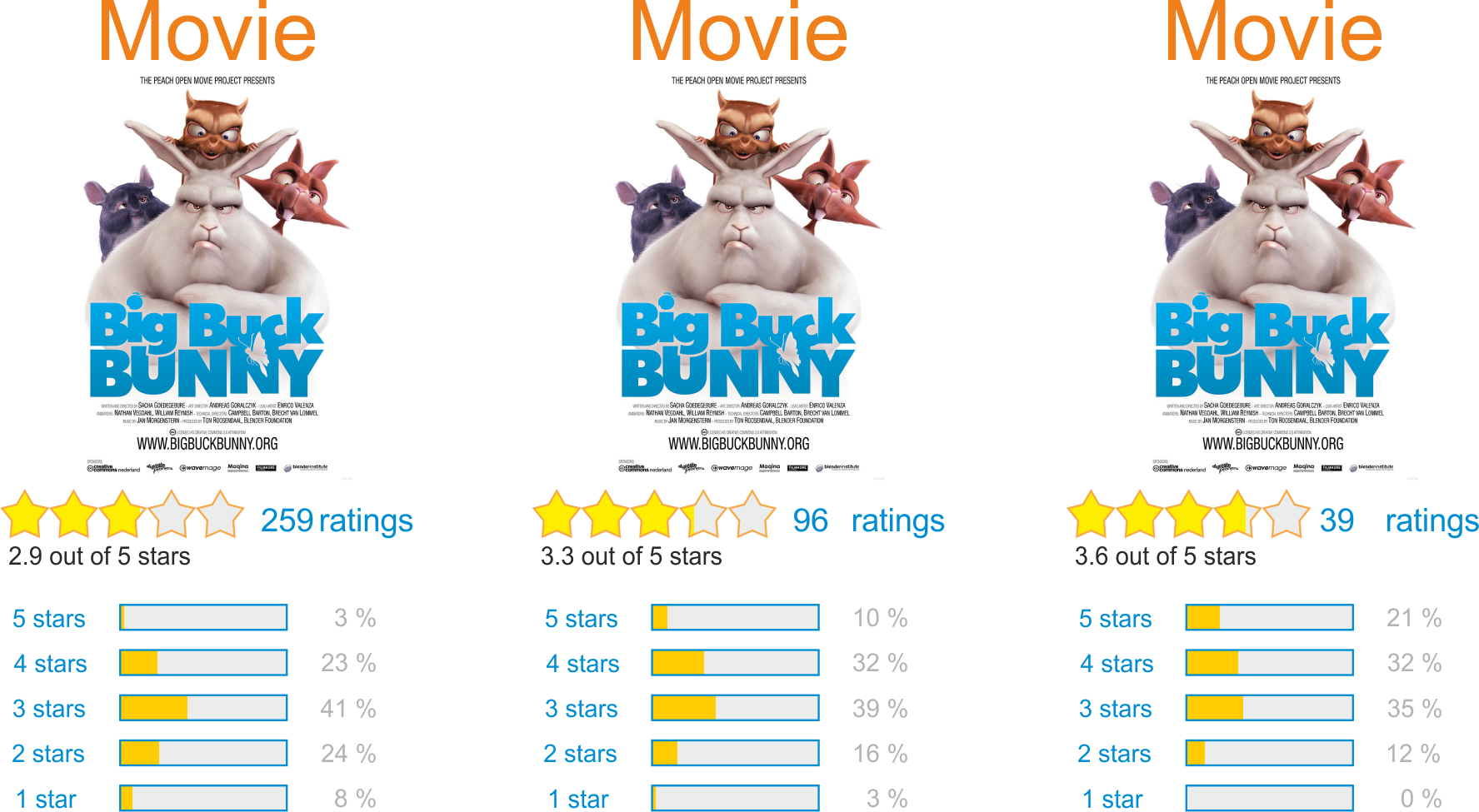}
\caption{An example snapshot of a choice set, with three different rating summary profiles based on different attribute levels. The movie poster was adopted from: \url{https://peach.blender.org/}.}
\label{fig:choice_set}
\end{figure*}

Conjoint choice experiments require a set of profiles, and a design how profiles are distributed into a number of choice sets. 

The identified attribute levels allow us to build a \textit{Full-Factorial design}~\cite{Zwerina1996ADesigns}, that consists of all possible combinations of attributes and levels, thus 2 attributes $\times$ 3 levels each result in 9 different profiles.
All profiles represent statistically feasible level combinations, while, for instance, a mean rating of 5 with a variance different from 0 would obviously be unfeasible.

In order to build the choice sets, and draw the most information on the interaction and main effects, three principles needed to be respected: \textit{level balance}, \textit{orthogonality} and \textit{minimal overlap}~\cite{Zwerina1996ADesigns}.
Level balance requires attribute levels to appear with equal frequency in the different choice sets. Orthogonality ensures that main and interaction effects are uncorrelated; this is achieved by having all attribute levels vary independently of each other. Overlap among levels for an attribute (i.e., identical attribute values for two or more profiles within the same choice set) reduces the collected information.
We used the D-efficiency metric to measure the statistical effectiveness of our design~\cite{Johnson2013ConstructingForce}:

 \begin{equation}
 D-\text{efficiency} = 100 \times \frac{1}{N \times |(\bm{X_C}' \bm{X_C})^{-1}|^{1/p}}
 \label{eq:d_eff}
 \end{equation}

Where $N$ is the number of observations in the design, as before, $p$ is the number of parameters, and $X_C$ is the standardized orthogonal contrast coding of the matrix $\bm{X}$ \cite{Kuhfeld2010}.  In matrix $\bm{X}$ columns correspond to the levels of each attribute. Each $m$ rows of the matrix $\bm{X}$, Figure~\ref{fig:des_mat}, where a single row is a binary representation of a profile in a choice set ($\bm{X_n}$).

Coding is the process or replacing our design levels by the set of indicator or coded variables. For determining the efficiency of the design we used the standard orthogonal contrast coding as recommended by \cite{Zwerina1996ADesigns}. Please notice that the sum of squares of the column in a standard orthogonal coding matrix is equal to the number of levels (e.g. if $X$ has two levels, the sum of squares of the columns of $X_C$ is 2). 
Thus, if $\bm{X}$ is orthogonal and balanced $\bm{X_C}' \bm{X_C} = N \bm{I}$ where $\bm{I}$ is a $p\times p$ identity matrix. In this case, the denominator terms in Formula~\ref{eq:d_eff} cancel each other, thus the efficiency is $100\%$.

\begin{figure}

\[ 
X = 
\begin{array}{c@{}c}
\left[
    \begin{BMAT}[5pt]{ccc.ccc}{ccc.cc}
      0& 1& 0& 0& 1& 0 \\
      1& 0& 0& 0& 0& 1\\
      0& 0& 1& 1& 0& 0\\
      \vdots & \vdots & \vdots & \vdots & \vdots & \vdots \\
      \vdots & \vdots & \vdots & \vdots & \vdots & \vdots
    \end{BMAT}
\right] 
& 
\begin{array}{l}
  \\[-30mm] \rdelim\}{3}{1mm}[$X_1$]
\end{array} 
\\
\hspace{0pt} L_1 \hspace{4pt} L_2 \hspace{4pt} L_3 \hspace{6pt} L_1 \hspace{4pt} L_2 \hspace{4pt} L_3
\\[-1ex]
\hexbrace{1.5cm}{A_1}\hexbrace{1.5cm}{A_2}
\end{array}
\]

\caption{Design matrix $X$, represented in Nonorthogonal Less-Than-Full-Rank Binary or Indicator Coding. }
\label{fig:des_mat}
\end{figure}

We identified a CBC design consisting of $N=6$ choice sets with $m=3$ alternatives to be optimal due to attaining 100\% D-efficiency with minimal overlap, balanced frequency of levels, and orthogonality of effects.

\subsubsection{Study procedure}

\paragraph{Choice Based Conjoint survey}

All participants were presented with the following hypothetical situation:

\textit{``Assume that you find yourself in the situation that you need to make a choice between three movies to watch on a movie platform. 
These three movies are equally preferable to you with respect to all other movie information you have access to (title, plot, actors etc.).
Other users' ratings are aggregated and summarized by their number of ratings, the mean rating value and their distribution.
Therefore, we would like to know your choice, by solely considering these rating summary statistics.''}

After assessment of demographics and choices on the origin of ratings, participants had to complete six choice tasks according to our design. Figure~\ref{fig:choice_set} depicts an exemplary choice set from the item-style run. The order of the choice tasks and the three answer options (i.e. profiles) were randomized for each respondent. 

\paragraph{Manipulation Check}
This task of the survey was again completed with a \textit{manipulation check} to validate  respondents' correct perception of our stimuli. 
In that manipulation check, we asked participants about the strategy they had employed in the making of their choices. 
Based on their answers, we only included those participants who reportedly had noticed the mean rating values, number of ratings or other aspects of the rating distribution. Those participants who claimed to have solely relied on their gut feelings, were removed from further analyses.

\subsubsection{Statistical analysis}

One of the basic assumptions underlying the assessment of users' choices is an additive utility model assuming that the different attributes and characteristics of an item/profile contribute, independently of each other, to the overall utility.
When confronted with a set of choices, respondents are supposed to select the alternative with the, in their eyes, maximal \textit{utility} $u$~\cite{Zwerina1996ADesigns}. The perceived utility of an item/profile is determined as:
\begin{equation}
\label{eq:utility}
u = \bm{x_i}\bm{\beta} + \epsilon
\end{equation}

where $\bm{x_i}$ is a vector characterizing a profile $i$, $\bm{\beta}$ is the vector with the unknown preferences for each attribute level, and $\epsilon$ is the residual error. 
 
The most common approach in analyzing CBC is the \textit{multinomial logit} ~\cite{Hauber2016StatisticalForce,Rao2014ChoiceAnalysis,Zwerina1996ADesigns}, where - given $N$ choice sets - each consisting of $m$ profiles, the probability of choosing profile $i$ in the choice set $n$ is defined by Equation~\ref{eq:probability_of_choice}:

\begin{equation}
\label{eq:probability_of_choice}
P(\text{choice}_n = i) = \frac{e^{\bm{x_{i}}\bm{\beta}}}{\sum_{j=1}^m e^{\bm{x_{j}}\bm{\beta}}}
\end{equation}

The multinomial logit is based on the assumption that the error $\epsilon$ is independent and identically distributed in a choice set.
We use the multinomial logit to estimate the coefficients of vector $\bm{\beta}$ that maximize the likelihood of a profile to be chosen based on respondents' data. 
The maximum likelihood estimator is consistent and asymptotically normal with the covariance matrix:
 \begin{equation}
 \Sigma = (Z'PZ)^{-1} = \big[ \sum_{n=1}^{N}\sum_{j=1}^{A} z'_{jn} P_{jn} z_{jn}\big]^{-1}
 \end{equation}

 \begin{equation}
 \text{where } z_{jn} = x_{jn} - \sum_{i=1}^{A} x_{in}P_{in}
 \end{equation}

Where the $\beta$ are the unknown parameters to be estimated.

\section{Results}
\label{sec:results}

Between January and February 2018 a group of \textit{77 people} participated in our choice experiments. In a smaller pre-study we validated the clarity and understandability of the questions and tasks. 
A total of 27 female and 50 male participants, whose age varied from 21 to 36, enrolled in a bachelor, or master or PhD in Computer Science, and all were familiar with online shopping and recommendation scenarios, participated in the main study. 23 respondents run the  item-style design, while, the remaining 54 were exposed to the user-style design. We removed 5 participants from further analyses, who failed the manipulation check. Thus, results are based on \textit{$72 \times 3 = 216$} choices with different origins of ratings (task 1) and \textit{$72 \times 6 = 432$} choices with different levels of number of ratings and mean values (task 2). The second task was alternated between participants (i.e. $50 \times 6 = 300$ were working on the user-style attribute levels and $22 \times 6 = 132$ on the item-style levels).

\subsection{Results on the origin of ratings}
In this Section, we present the obtained results from the controlled experiment on the origin of ratings. 
Given the entire set of observations from the 3 choice sets we measured the frequencies with which each alternative was chosen. We present results in Table~\ref{tab:origin_results}. The preferred source of ratings are the similar users (selected 95 times), own ratings on similar items is slightly less preferred (selected 89 times). Whereas, clearly fewer respondents preferred items supported by connections on social network (selected 32 times).
Furthermore, we analyzed the responses from the manipulation check, which was presented as a multiple select question. We noticed that the similar user justification and own ratings on similar items was checked 54 times and 46 times respectively, while the connections on social networks (like Facebook) were only considered 10 times as one of the primary decision heuristics for this set of questions. 

This can be related to the fact that a social network link does not necessarily mean friendship in the everyday sense, and there can be several reasons, beyond friendship, for people to connect \cite{boyd2007SocialScholarship}. While respondents tend to trust analog behaviors to similar users and past choices.

\begin{table}[hb]
\caption{Frequency count for the origin of ratings in the compared choice sets and the manipulation check for the item style.}
\begin{tabular}{lcc}
\hline
Origin & Freq. in Choice Sets & Freq. in Manipulation check\\
\hline
Similar users & 95 & 54\\
Similar items & 89 & 46\\
Facebook Friends & 32 & 10\\
\hline
\end{tabular}
\label{tab:origin_results}
\end{table}

\subsection{Multinomial Logit results}
\begin{table*}[th!]
\caption{Final results  of the multinomial logit model for the user style.}
\centering

\begin{tabular}{lllcccc}
\toprule
Style & Attribute & Lvl.  & Lvl. value&Coefficient & Standard Error & P Value\\
\midrule
\multirow{6}{*}{\shortstack[l]{Item style}} & \multirow{3}{*}{\shortstack[l]{Number of Ratings}} & Large & 259& 1.25	& 0.25		 & $<0.0001$ \\
								&& Medium & 96& 0.74 &	0.27		& 0.0063\\
                            	&& Small & 39&\multicolumn{3}{l}{Constrained to be 0}\\

\cline{2-7}

&\multirow{3}{*}{Mean Rating} & High &  4& 2.07	& 0.31	&	$<0.0001$	 \\
&								& Average & 3.7 &  0.73	& 0.35	&	0.04 \\
&                            	& Low & 3.4&\multicolumn{3}{l}{Constrained to be 0} \\         
\midrule
\multirow{6}{*}{\shortstack[l]{User style}} & \multirow{3}{*}{\shortstack[l]{Number of Ratings}} & Large &2970& 0.53	&0.14		 & 0.0002 \\
								&& Medium &560&  0.13 &	0.16		& 0.39 \\
                            	&& Small & 290&\multicolumn{3}{l}{Constrained to be 0}\\

\cline{2-7}

&\multirow{3}{*}{Mean Rating} & High &3.6&  2.77	& 0.27	&	$<0.0001$	 \\
&								& Average &3.3 &  1.09	& 0.29		&	0.0002 \\
&                            	& Low & 2.9& \multicolumn{3}{l}{Constrained to be 0} \\

\bottomrule
\end{tabular}
\label{tab:resultsmultilogit}
\end{table*}

In this subsection, we present the obtained results from the two runs of the CBC. For the entire sample of observations, we estimated the multinomial logit model underlying the CBC design. 

Detailed results for estimating the preference weights for the item-style and the user-style are presented in Table~\ref{tab:resultsmultilogit}. The fourth column of Table~\ref{tab:resultsmultilogit} shows the coefficients (or preference weights) for each level of the two attributes and two styles, where the base levels (i.e. small number of ratings and low mean ratings) have been constrained to be zero. The fifth and the sixth column of Table~\ref{tab:resultsmultilogit} report the standard errors and the P-values for the respective levels of each attribute.

Figure~\ref{fig:multiLogit1} and \ref{fig:multiLogit2} visually depict the preference weights of the multinomial logit model for each level of the two selected attributes (i.e. total number and mean of ratings) for the item-style and the user-style attribute levels correspondingly. As expected, there was a general {\it higher is better} tendency for the two attributes - i.e., users prefer bigger numbers of ratings and higher mean values. 

However, for the user-style attribute levels, there is a clear and statistically significant preference relation over the three levels for mean rating values. However, in terms of the total number of ratings, users do not seem to care that much. The large number of ratings is statistically significant, and clearly preferred over the other two levels, but between the medium and small level, the P-value is above the threshold of .05 (cmp. Table~\ref{tab:resultsmultilogit}) and thus no statistically noticeable difference in the users' perception exists.

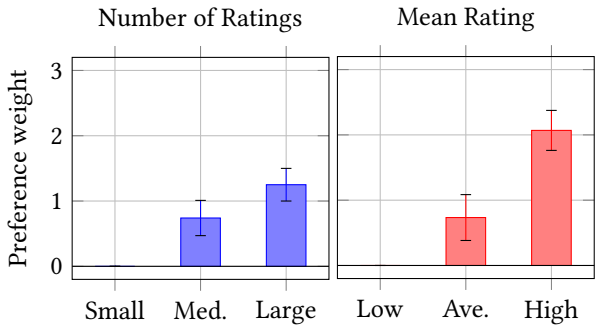
\begin{figure}
\centering
\begin{subfigure}{.30\textwidth}
\begin{tikzpicture}
\begin{axis}[
    title = {Number of Ratings},
    ylabel= Preference weight,
    width=5cm,
    height = 4.5cm,
    ymin=-0.2,
    ymax = 3.2,
    bar width=15pt,
    enlarge x limits={abs=0.5},
    xtick={1,...,3},
    xticklabels={%
        Small,
        Med.,
        Large},
    grid=major,
    ybar
    ]

\addplot[
    fill=blue!50,
    draw=blue,
    point meta=y,
    every node near coord/.style={inner ysep=5pt},
    error bars/.cd,
        y dir=both,
        y explicit
] 
table [y error=error] {
x   y           error    label
1   0   		0 			0 
2   0.74	0.27 1
3   1.25	0.25 2 
};

\draw ({rel axis cs:0,0}|-{axis cs:0,0}) -- ({rel axis cs:1,0}|-{axis cs:0,0});
\end{axis}
\end{tikzpicture}
\end{subfigure}
\begin{subfigure}{.30\textwidth}
\begin{tikzpicture}
\begin{axis}[
    title = {Mean Rating},
    width=5cm,
    height = 4.5cm,
    ymin=-0.2,
    ymax = 3.2,
    xtick={1,...,3},
    yticklabels={,,},
    bar width=15pt,
    enlarge x limits={abs=0.5},
    xticklabels={%
        Low,
        Ave.,
        High},
    grid=major,
    ybar
    ]

\addplot[
    fill=red!50,
    draw=red,
    point meta=y,
    every node near coord/.style={inner ysep=5pt},
    error bars/.cd,
        y dir=both,
        y explicit
] 
table [y error=error] {
x   y           error    label
1   0   		0 			0 
2   0.73412	0.35118	 1
3  2.06892	0.30636	 2 
};

\draw ({rel axis cs:0,0}|-{axis cs:0,0}) -- ({rel axis cs:1,0}|-{axis cs:0,0});
\end{axis}
\end{tikzpicture}
\end{subfigure}

\caption{Preference weights of the multinomial logit model on the item style run.}\label{fig:multiLogit1}

\end{figure}
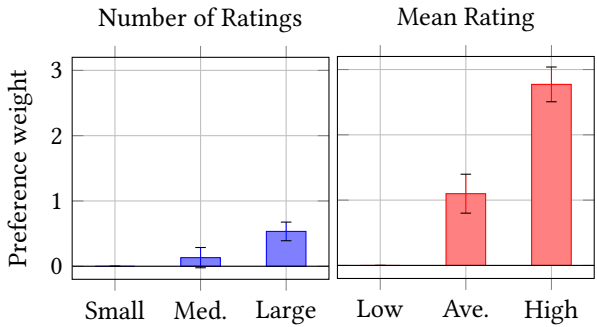
\begin{figure}

\centering
\begin{subfigure}{.30\textwidth}
\begin{tikzpicture}
\begin{axis}[
    title = {Number of Ratings},
    ylabel= Preference weight,
    width=5cm,
    height = 4.5cm,
    ymin=-0.2,
    ymax = 3.2,
    bar width=15pt,
    enlarge x limits={abs=0.5},
    xtick={1,...,3},
    xticklabels={%
        Small,
        Med.,
        Large},
    grid=major,
    ybar
    ]

\addplot[
    fill=blue!50,
    draw=blue,
    point meta=y,
    every node near coord/.style={inner ysep=5pt},
    error bars/.cd,
        y dir=both,
        y explicit
] 
table [y error=error] {
x   y           error    label
1   0   		0 			0 
2   0.13193	0.15510 1
3   0.53364	0.14262 2 
};

\draw ({rel axis cs:0,0}|-{axis cs:0,0}) -- ({rel axis cs:1,0}|-{axis cs:0,0});
\end{axis}
\end{tikzpicture}
\end{subfigure}
\begin{subfigure}{.30\textwidth}
\begin{tikzpicture}
\begin{axis}[
    title = {Mean Rating},
    width=5cm,
    height = 4.5cm,
    ymin=-0.2,
    ymax = 3.2,
    xtick={1,...,3},
    yticklabels={,,},
    bar width=15pt,
    enlarge x limits={abs=0.5},
    xticklabels={%
        Low,
        Ave.,
        High},
    grid=major,
    ybar
    ]

\addplot[
    fill=red!50,
    draw=red,
    point meta=y,
    every node near coord/.style={inner ysep=5pt},
    error bars/.cd,
        y dir=both,
        y explicit
] 
table [y error=error] {
x   y           error    label
1   0   		0 			0 
2   1.09860	0.29814	 1
3  2.77258	0.26614 2 
};

\draw ({rel axis cs:0,0}|-{axis cs:0,0}) -- ({rel axis cs:1,0}|-{axis cs:0,0});
\end{axis}
\end{tikzpicture}
\end{subfigure}

\caption{Preference weights of the multinomial logit model on the user style run.}\label{fig:multiLogit2}
\end{figure}

While, for the item-style, it is obvious that users also prefer the high mean rating, but the preference weights on the higher numbers of ratings are clearly higher than for the user-style attribute levels. This can be explained by the lower levels for the number of ratings attribute in the items-style setup, since average ratings per user and thus the ratings of similar items from the current user are obviously lower than the average number of ratings for items. According to our findings users are more sensitive to changes of the number of ratings on a smaller scale (i.e. two digit and low three digit numbers) rather than changes on a larger scale (i.e. up to 4 digit number of ratings). 


\begin{table}
\caption{Probability of choice over profiles from the item style run, in decreasing order.}
\centering
\begin{tabular}{lllcc}
\hline
 &\# of Ratings & Mean Rating & Pr. of choice & Utility\\
\hline
1	&	Large	&	High	&	38.18	\%	&	3.32	\\
2	&	Medium	&	High	&	22.93	\%	&	2.81	\\
3	&	Small	&	High	&	10.94	\%	&	2.07	\\
4	&	Large	&	Average	&	9.99	\%	&	1.98	\\
5	&	Medium	&	Average	&	6.01	\%	&	1.47	\\
6	&	Small	&	Average	&	4.82	\%	&	1.25	\\
7	&	Large	&	Low	&	2.89	\%	&	0.74	\\
8	&	Medium	&	Low	&	2.86	\%	&	0.73	\\
9	&	Small	&	Low	&	1.38	\%	&	0.00	\\
\hline

\end{tabular}
\label{tab:prChoice1}
\end{table}

\begin{table}
\caption{Probability of choice over profiles from the user style run,  in decreasing order.}
\centering
\begin{tabular}{lllcc}
\hline
 &\# of Ratings & Mean Rating & Pr. of choice & Utility\\
\hline
1	&	Large	&	High	&	35.47	\%	&	3.31	\\
2	&	Medium	&	High	&	23.73	\%	&	2.90	\\
3	&	Small	&	High	&	20.80	\%	&	2.77	\\
4	&	Large	&	Average	&	6.65	\%	&	1.63	\\
5	&	Medium	&	Average	&	4.45	\%	&	1.23	\\
6	&	Small	&	Average	&	3.90	\%	&	1.10	\\
7	&	Large	&	Low	&	2.22	\%	&	0.53	\\
8	&	Medium	&	Low	&	1.48	\%	&	0.13	\\
9	&	Small	&	Low	&	1.30	\%	&	0.00	\\
\hline

\end{tabular}
\label{tab:prChoice2}
\end{table}


From the different levels of preference weights (partial utilities) for the two signals (i.e. levels of the profile attributes) we can also derive the perceived overall utility (see Table~\ref{tab:prChoice1} and Table~\ref{tab:prChoice2}). Following formula \ref{eq:probability_of_choice} the probability of selecting any of the 9 profiles was computed and ordered by decreasing values in Table~\ref{tab:prChoice1} for the item-style, and Table~\ref{tab:prChoice2} for the user-style attribute levels. 

For the item-style attribute levels, the two highest levels of the mean value were well perceived, instead the small number or ratings with high mean signified almost similar utility to users like the large number of interactions with a slightly lower (average) mean value, cmp.Table VI utility $2.07$ vs. $1.98$.
Whereas in the user-style run, changes in the mean value were strongly perceived, while changes in the relatively high numbers of ratings had far less impact on users' choice - i.e., an increase in the mean rating value by one level increased the probability of choice by a factor of three to four, when everything else was kept constant.

\section{Discussion}
\label{sec:discussion}
Rating summarizations provide important clues to users in online choice situations. Marketing research has shown that consumers are strongly guided by online reviews, and that the mean rating value is interpreted as an indicator for the quality of a product~\cite{duan2008online}. Also in our study, participants seem to have been following this quality hypothesis. However, research comparing online reviews with scientific product testing, identified that the average star rating has a surprisingly low correspondence to established quality metrics~\cite{HBR_Ratings}.

The total number of ratings, on the other hand, is typically regarded as an indicator for the popularity of a product or an item in general. Given that with larger sample sizes, all things being equal, the mean rating value becomes more informative, it is also very reasonable that, in case of a large number of ratings, users would be more likely to follow this choice. 

This work is in line with prior research on the effects of potential decision biases such as position, decoy or framing effects, on the choice behavior of users \cite{chen2013human,teppan2015decision}. Implications of these findings about the inherent dynamics of online choice behavior can be either purposefully exploited to develop more persuasive systems \cite{yoo2012persuasive} or explicitly neutralized, as proposed by \cite{teppan2012minimization}.   

In the line of exploitation of such effects, one can see the work of Adollahi and Nasraoui~\cite{Abdollahi2017UsingFactorization}, who extended Matrix Factorization with a soft constraint that brings explainable items closer to users in the latent space. Their concept of {\it explainability} builds on user-style explanations \cite{Herlocker2000ExplainingRecommendations}, where rating summary statistics from a users' nearest neighbors are presented. Abdollahi and Nasraoui consider an item only to be explainable, if the mean rating value within this neighborhood is beyond a threshold. Thus, items with higher mean rating values within a user's neighborhood are pushed towards higher ranks in recommendation lists, which - according to our empirical results - should also increase their probability of being selected. In future research using this experimental paradigm we intend to explore the impact of additional parameters describing rating summary statistics, like the variance and skewness that we controlled in this study with fixed values, corresponding to the respective medians of the Netflix dataset. Furthermore, we will comprehensively exploit these parameters in algorithm development.

\section{Conclusions}
In this paper we conducted a user study to explore two aspects of explanations: how users react to different origins of rating, and how they perceive rating summary statistics in choice setting. 
As far as the origin of ratings is concerned, we noticed that users are reluctant to recommendations justified with links on social networks, while recommendations justified with user-style and item-style collaborative explanations were significantly more preferred.
Furthermore, we explored how the total number and the mean of ratings influenced users' choices. Results clearly demonstrate that choice behavior was heavily influenced by the mean rating value, while a big number of ratings only modestly influences the choices users make. However, when comparing smaller levels of rating numbers users are stronger influenced by an increase from a two digit to a three digit number of ratings. This study used representative attribute levels for ratings in the movie domain and shed additional light on the influence of rating summary statistics on users' choice behavior. Future work will focus on extending the research to additional attribute levels and on developing algorithms that foster items with higher choice probability according to these findings.

\newpage
\bibliographystyle{ACM-Reference-Format}
\bibliography{biblio,Ludovik_Mendeley}

\end{document}